\documentclass{article}
\usepackage[top=3cm, bottom=3cm, left=3cm, right=3cm]{geometry}
\usepackage{amsmath}
\usepackage{amsfonts}
\usepackage{amssymb}
\usepackage{graphicx}
\begin{document}

\begin{center}
\begin{large} \textbf{Ground-State Masses and Magnetic Moments of Heavy Baryons}
\end{large}

\vspace{1cm}

Zahra Ghalenovi$^{a,b}$, Ali Akbar Rajabi$^{b}$, Si-xue Qin $^{a}$, Dirk. H. Rischke$^{a,c}$ \\ 
\vspace{0.3cm}
\small{

$^{a}$Institute for Theoretical Physics, Johann Wolfgang Goethe University, Frankfurt am Main, Germany\\
$^{b}$Physics Department, Shahrood University of Technology, Shahrood, Iran}\\
$^{c}$Frankfurt Institute for Advanced Studies, Frankfurt am Main, Germany\\

\end{center}

\vspace{0.5cm}

\begin{abstract}
\noindent  In this work we study single, double, and triple
heavy-flavor baryons using the hypercentral approach in the framework
of the non-relativistic quark model. Considering two different
confining potentials and an improved form of the hyperfine
interaction, we calculate the ground-state masses of heavy baryons and
also the ground-state magnetic moments of single charm and beauty
baryons with $J^P =3/2^+$. The obtained results are in good agreement 
with experimental data and those of other works.
\end{abstract}

\section{Introduction}

The properties of heavy-flavor baryons have recently received much
attention, both experimentally and theoretically
\cite{Albertus:2006ya, Guo:1996eg, Ebert:2002ig, Zhang:2008pm}. 
The investigation of the properties of such hadrons is not only important
to understand the dynamics of quantum chromodynamics (QCD) at hadronic
energy scales, but also interesting in view of the recent progress in
studying heavy-flavor hadrons by different experimental groups 
like BaBar, BELLE, BESIII, CLEO, and SELEX. Different methods based on
the constituent quark model (CQM) have been used to investigate heavy-flavor baryons. 
Ebert et al.\ studied heavy baryons in the quark-diquark model in the 
relativistic limit \cite{Ebert:2005xj}. Reference \cite{Guo:1996eg}
investigated heavy-flavor baryons by using the Bethe-Salpeter equation in
the heavy-quark limit and calculated the Isgur-Wise function. 
Albertus et al.\ evaluated different properties of single heavy-flavor
baryons using heavy-quark symmetry in the non-relativistic quark model
\cite{Albertus:2005uq}. Flynn et al.\ studied charmed baryons and 
spin-splittings in quenched lattice QCD \cite{Flynn:2003vz}. Faessler
et al.\ considered ground-state magnetic moments of heavy baryons in 
the relativistic quark model using heavy-hadron chiral perturbation
theory \cite{Faessler:2006ft}. Patel et al.\ used the non-relativistic
quark model with a hypercentral Coulomb plus linear potential and 
obtained masses and magnetic moments of heavy-flavor baryons 
\cite{Patel:2007gx, Patel:2008mv}. 

In the present work we calculate the ground-state masses and magnetic
moments of heavy baryons in the hypercentral approach 
\cite{Ghalenovi:2011zz, Isgur:1978xj, Capstick:1986bm,
  Giannini:1990pc, Glozman:1995fu, Salehi:2011zza, Hassanabadi:2008zz,
  Hassanabadi:2005z, Rajabi:2005}. 
We study the three-body problem, particularly the baryons containing
one, two, and three charm (beauty) quarks. The potential is assumed to
be a combination of a long-range confinement part and a short-range
potential which is a Coulombic one, depending on the color charge.

The solution of a three-body problem in three spatial dimensions 
is rather difficult. Here, we employ the hypercentral approach where 
the Schr\"{o}dinger equation of the three-body system depends only on
a single variable. We solve this one-dimensional Schr\"{o}dinger
equation numerically. We also introduce a non-confining interquark
potential, namely a spin-isospin dependent part, as hyperfine interaction. 

We study the baryonic systems using two types of potentials. First, we 
introduce the Cornell potential, $bx-c/x$, as confining potential
between quarks and obtain the masses of heavy baryons. Second, we add
a harmonic oscillator term to the confining potential and then
compare the obtained baryon masses to the results without this term,
and also to those of other works. The obtained masses and magnetic
moments are close to experimental data and other theoretical predictions.

This paper is organized as follows. In Sec.\ 2 we introduce the 
interquark potential. In Sec.\ 3 we simplify the three-body problem 
using the hypercentral approach. We present our method to obtain
masses and magnetic moments of baryons in Sec.\ 4. Numerical results 
are shown and compared to those of other works in Sec.\
5. Finally, a summary is given in Sec.\ 6.

\section{Interaction Potential}
In principle, the potential between quarks could be of any confining
form (e.g.\ linear, logarithmic, power law, etc.). The interquark
potential usually contains a linear part which describes confinement
in QCD and is supplemented by a Coulomb term which may be attributed 
to one-gluon exchange. The Coulomb term alone is not sufficient 
because it would allow ionization of quarks from the system. 
As a first case (in the following termed ``case I''), we consider the 
Cornell potential \cite{Bali:2000vr, Guillaume:2005v}:
\begin{equation}\label{1}
V(x)=bx-\frac{c}{x}\;,
\end{equation}
where $x$  is the relative coordinate of the quark pair, 
and $b,\,c$ are constants. In many practical applications a 
harmonic oscillator (h.o.) potential yields spectra not much different
from those for Eq.\ (1) \cite{Bali:2000vr}. Therefore, as a second
case (termed ``case II'') we also consider a potential which is a
combination of Eq.\ (1) and the h.o.\ potential which has the form $ax^{2}$ :
\begin{equation}\label{2}
V(x)=ax^2+bx-\frac{c}{x}\;,
\end{equation}
where $a$ is another constant. 
In addition, we introduce a spin- and isospin-dependent potential as 
hyperfine interaction for the baryons. This combination of potentials 
yields spectra which are very close to the experimental results and
other theoretical predictions. 

The non-confining spin-spin interaction potential is proportional 
to a $\delta$-function which is an illegal operator term 
\cite{Giannini:2003xx}. We modify it to a Gaussian function of the
relative distance of
the quark pair,
\begin{equation}\label{3}
H_{S}=A_{S}\,\frac{\vec{s}_1 \cdot \vec{s}_2}{(\sqrt {\pi}\sigma_S)^3}\,
\exp\left(-\frac{x^2}{\sigma_S^2}\right)\;,
\end{equation}
where $s_{i}$ is the spin operator of the $i^{th}$ quark ($\vec{s}_i =
\vec{\sigma}_i/2$, with $\vec{\sigma}_i$ being the vector of Pauli
matrices) and $A_{S}$ and $\sigma_{S}$ are constants. 

Other spin-, as well as isospin-dependent interaction potentials can
arise from quark-exchange interactions.
We conclude that two additional terms should be added to the
Hamiltonian for quark pairs which result in hyperfine interactions 
similar to Eq.\ (3). The first one depends on isospin only and 
has the form \cite{Giannini:2003xx, Ghalenovi:2012zz}:
\begin{equation}\label{4}
H_{I} =A_{I}\,\frac{\vec{t}_1 \cdot \vec{t}_2}{(\sqrt {\pi}\sigma_I)^3}\,
\exp\left(-\frac{x^2}{\sigma_I^2}\right)\;,
\end{equation}
where $t_{i}$ is the isospin operator of the $i^{th}$ quarks, 
and $A_{I}$ and $\sigma_{I}$ are constants.
The second one is a spin-isospin interaction given by \cite{Giannini:2003xx, Ghalenovi:2012zz}:
\begin{equation}\label{5}
H_{SI} =A_{SI}\,\frac{(\vec{s}_1 \cdot
\vec{s}_2)(\vec{t}_1 \cdot \vec{t}_2)}{(\sqrt {\pi}\sigma_{SI})^3}\,
\exp\left(-\frac{x^2}{\sigma_{SI}^2}\right)\;,
\end{equation}
where $s_{i}$ and $t_{i}$ are the spin and isospin operators 
of the $i^{th}$ quark, respectively, and $A_{SI}$ and $\sigma_{SI}$
are constants. Then, from Eqs.\ (3-5) 
the hyperfine interaction (a non-confining potential) is given by
\begin{equation}\label{6}
H_{int}(x) = H_{S}(x)+H_{I}(x) + H_{SI}(x)\;.
\end{equation}
The parameters of the hyperfine interaction (6) are given in Table 1.

\begin{table}
\begin{center}
Table 1. Constituent quark masses and hy\-per\-fine - po\-ten\-tial \\parameters used in cases I and II \cite{Ghalenovi:2011zz, Ghalenovi:2012iu}.\\
{\begin{tabular}{c c} \hline  \hline  
parameter                   & value           \\
\hline 
$m_{u}$                     &  330   MeV      \\ 
$m_{d}$                     &  335   MeV      \\
$m_{s}$                     &  469   MeV      \\
$m_{c}$                     &  1600  MeV      \\
$m_{b}$                     &  4980  MeV      \\
$\sigma_{S}$                &  2.87  fm     \\
$A_{S}$                     &  67.4  fm$^{2}$ \\
$\sigma_{SI}$               & 2.31   fm     \\
$A_{SI}$                    & -106.2 fm$^{2}$ \\
$\sigma_{I}$                & 3.45   fm     \\
$A_{I}$                     &51.7    fm$^{2}$\\
\hline  \hline  

\end{tabular}}
\end{center}
\end{table}

\section{The Hypercentral Approach}

In the quark model, a baryon is a three-body bound state made of
quarks. The mathematical description of a three-body system is
more complicated than that of a two-body system. 
Several methods have been used by different authors to solve 
three-body problems 
\cite{Hassanabadi:2008zz,Hassanabadi:2005z,Rajabi:2005,Giannini:2003xx,Ghalenovi:2012zz,Patel:2002jj,Ghalenovi:2012iu,Patel:2008nv}.

In order to describe the baryon as a bound state of three constituent quarks, 
we define the configuration of three particles by two 
Jacobi coordinates $ \rho $ and $ \lambda $  as 
\begin{equation}\label{7}
\vec{\rho}=\frac{1}{\sqrt{2}}(\vec{r}_1-\vec{r}_2)\;,  
\qquad\vec{\lambda}=\frac{1}{\sqrt{6}}(\vec{r}_1+\vec{r}_2-2\vec{r}_3)\;,
\end{equation}
such that
\begin{equation}\label{8}
m_{\rho}=\frac{2m_{1}m_{2}}{m_{1}m_{2}}\;,
\qquad m_{\lambda}=\frac{3m_{3}(m_{1}+m_{2})}{2(m_{1}+m_{2}+m_{3})}\;.
\end{equation}
Here  $ m_{1} $ ,  $ m_{2} $, and  $ m_{3} $  are the constituent
quark masses.  Instead of $ \rho $ and $ \lambda $ , one can 
introduce hyperspherical coordinates which are given by the 
angles $\Omega_\rho=(\theta_\rho,\phi_\rho)$ and  
$\Omega_\lambda=(\theta_\lambda,\varphi_\lambda)$, respectively,  
together with the hyperradius $ x $  and the hyperangle  $ \zeta $ , defined by
\begin{equation}\label{9}
x=\sqrt{\vec{\rho}\,^2+\vec{\lambda}\,^2}\;,
\qquad \zeta=\arctan\left(\frac{\rho}{\lambda}\right)\;.
\end{equation} 
Therefore, the Hamiltonian will be
\begin{equation}\label{10}
H=\frac{P_\rho^2}{2m_{\rho}}+\frac{P_\lambda^2}{2m_{\lambda}}+V(\rho,\lambda)
=\frac{P^2}{2m}+V(x)\;.
\end{equation}
In the hypercentral constituent quark model (hCQM), the quark
potential, $ V $, is assumed to depend only on the hyperradius $x$. 
Therefore, in the three-quark wave function one can factor out the
hyperangular part which is given by hyperspherical harmonics. 
The remaining hyperradial part of the wave function is determined by
the hypercentral Schr\"{o}dinger equation  
\begin{equation}\label{11}
\left[\frac{d^2}{dx^2}+\frac{5}{x}\frac{d}{dx}
-\frac{\gamma(\gamma+4)}{x^2}\right]\psi_\gamma(x)=-2m[E_\gamma-V(x)]\psi_\gamma(x)\;,
\end{equation}
where $ \psi_\gamma(x) $ , $ E_\gamma $, and $ \gamma $  are the 
hyperradial part of the wave function, the energy eigenvalues, and the
grand angular quantum number, respectively. The latter is given by 
$ \gamma=2\nu+l_{\rho}+l_{\lambda} $ where $ l_{\rho} $ and $
l_{\lambda} $  are the angular momenta associated with the $ \rho $  
and $ \lambda $  variables and  $ \nu $ is a non-negative integer
number. The quantity $ m $  in Eqs.\ (10,11) is the reduced mass, 
\begin{equation}\label{12}
m=\frac{2m_{\rho} m_{\lambda}}{m_{\rho}+m_{\lambda}}\;.
\end{equation}
We use the transformation
\begin{equation}\label{13}
\psi_{\gamma}(x)=x^{-5/2}\phi_{\gamma}(x)
\end{equation}
to bring Eq.\ (11) into the form
\begin{equation}\label{14}
\left[\phi^{\prime\prime}_{\gamma}(x)
-\dfrac{(2\gamma+3)(2\gamma+5)}{4x^{2}}\right] \phi_{\gamma}(x)
=-2m[E_{\gamma}-V(x)]\phi_{\gamma}(x)\;.
\end{equation}
Substituting the potentials (1) and (2) into Eq.\ (11) we obtain 
the following equations:
\begin{enumerate}
\item[(I)] In case I we only consider the Cornell potential (1) as 
confining interaction. Using the hyperradial approximation 
used in Ref.\ \cite{Ghalenovi:2011zz}, the Schr\"{o}dinger equation 
for the baryons is given as
\begin{equation}\label{15}
\phi^{\prime\prime}_{\gamma}(x)+2\mu\left[-bx+\dfrac{c}{x}
-\dfrac{(2\gamma+3)(2\gamma+5)}{8 \mu x^{2}}\right]\phi_{\gamma}(x)
=-2\mu E_{\gamma}\phi_{\gamma}(x)\;,
\end{equation}
where $ \mu =m $. As in Ref.\ \cite{Ghalenovi:2011zz}, 
in the following we shall consider  $ \mu$ as a free parameter 
which is fitted to the baryon spectrum.
\item[(II)] In case II, we add the h.o.\ term to the confining
  interaction. Then, using the potential (2) and substituting it into 
Eq.\ (14) we obtain the following equation:
\begin{equation}\label{16}
\phi^{\prime\prime}_{\gamma}(x)+\left[ -a_{1}x^2-b_{1}x
+\frac{c_{1}}{x}-\dfrac{(2\gamma+3)(2\gamma+5)}{4x^{2}}\right]
\phi_{\gamma}(x)
=-\varepsilon_{\gamma}\phi_{\gamma}(x)\;,
\end{equation}
where
\begin{equation}\label{17} 
\varepsilon_{\gamma}=2mE_{\gamma}\;,\;\; a_{1}=2ma\; ,\;\; 
b_{1}=2mb \;,\;\;  c_{1}=2mc \;.
\end{equation}
\end{enumerate}

\section{Heavy Baryon Masses and Magnetic Moments }

We take the non-confining potential $H_{int}$, Eq.\ (6), as a
perturbation of the energy eigenvalues obtained by solving Eqs.\
(15,16). To first order in perturbation theory, the correction
can be computed using the unperturbed wave function for the ground state,
\begin{equation}\label{18} 
\langle H_{int}\rangle=\dfrac{\int\psi_{0}H_{int}\psi_{0}\,x^{5}dx\,
d\Omega_{\rho}\,d\Omega_\lambda}{\int\psi_{0}\psi_{0}\,x^{5} dx\,
d\Omega_{\rho}\,d\Omega_\lambda}\;.
\end{equation}
Note that in Eqs.\ (3-5) the spin-isospin dependent 
interaction potentials of a two-quark system are actually functions of the
relative distance between the quarks. In the hypercentral approach, 
however, we take the same form for these potentials, but
replace the relative distance by the hyperradius $x$ which is the average
relative distance between the three quarks. We believe that this is 
a reasonable approximation, at least for quarks with the same mass.
The spin-spin term $\vec{s}_1 \cdot \vec{s}_2$ in Eq.\ (3) is
replaced by the average of $\sum_{i<j}{\vec{s}_i \cdot \vec{s}_j}$,
and similarly for Eqs.\ (4,5).

The mass of the baryon is then obtained as the sum of the masses of
the constituent quarks, the ground-state energy eigenvalue $E_0$, and
$\langle H_{int}\rangle$,
\begin{equation}\label{20} 
M_{B}=m_{1}+m_{2}+m_{3}+E_{0}+\langle H_{int}\rangle\;,
\end{equation}
where $ E_{0}+\langle H_{int}\rangle $ depends on the 
type of confining interaction used.  
The effective quark mass is defined as 
\begin{equation}\label{21}
m_{i}^{eff}=m_{i}\left(1+\frac{E_0+\langle H_{int}\rangle}{\sum_{i}m_{i}}\right)\;,
\end{equation}
such that the mass of the baryon is
\begin{equation}\label{22}
M_{B}=\sum_{i}m_{i}^{eff}\;.
\end{equation}
The physical interpretation of the effective quark mass (\ref{21}) is
that, within a baryon, the mass of a quark may get modified due to 
its interactions with the other quarks. 

In the quark model, the magnetic moment of the baryon is 
obtained as \cite{Patel:2007gx, Patel:2002jj}
\begin{equation}\label{23}
{\mu}_{B}=\left\langle \phi_{sf}|M_z|\phi_{sf}\right\rangle\;,
\end{equation}
where $|\phi_{sf}\rangle$ represents the spin-flavor wave function of 
the respective baryonic state and
\begin{equation}\label{24}
\vec{M} = \sum_{i}\frac{g e_i \vec{s}_{i}}{2m_{i}^{eff}}\;.
\end{equation}
Here, $g=2$ is the spin $g-$factor, $ e_{i} $ is the electric charge, 
and $\vec{s}_{i}$ the spin of the $i^{th}$ quark.

\section{Discussion}

In Refs.\ \cite{Ghalenovi:2012zz, Ghalenovi:2012iu} heavy-flavor
baryons were studied in the hypercentral approach with the confining
interaction (2) and the hyperfine interactions (3-5). 
The Schr\"{o}dinger equation was solved analytically to obtain masses 
of heavy-flavor baryons. In Ref.\ \cite{Ghalenovi:2011zz} heavy-flavor
baryons were also studied in the hypercentral approach, but with the
confining interaction (1) and the hyperfine interactions (3-5). 
The Schr\"{o}dinger equation was solved using a variational method 
to obtain masses of single, double, and triple heavy-flavor baryons.
Patel et al.\ used the non-relativistic quark model with hypercentral 
Coulomb plus linear potential  \cite{Patel:2007gx, Patel:2002jj} and 
Coulomb plus harmonic oscillator potential \cite{Patel:2008nv} and 
obtained heavy-flavor baryon masses. 

\begin{center}
Table 2. The confining potential parameters  \\ 
used in case I and II \cite{Ghalenovi:2011zz, Ghalenovi:2012iu}.\\
\begin{tabular}             {l l l} \hline  \hline  
 &  parameter        &     value          \\
\hline 
    caseI     &b         &     1.61    $fm^{-2}$ \\
 
      &   c       &     4.59               \\
      \\
  & a                &     0.73    $fm^{-3}$    \\
caseII   &  b        &     0.81    $fm^{-2}$     \\
     &      c    &     2.12                   \\
\hline  \hline  
\end{tabular}

\end{center}

In the present work we use the same potentials and the hypercentral
approach, but we solve the Schr\"{o}dinger equation numerically. 
We study baryonic systems using the confining interactions (1) and
(2), respectively.  The quark masses and potential parameters used in 
both case I and II are obtained from our corresponding works 
\cite{Ghalenovi:2011zz} and \cite{Ghalenovi:2012iu}, respectively, 
and are listed in Table 1. The confining potential parameters for the 
two cases are listed in Table 2. In case I, the parameter $ \mu $  of
Eq.\ (15) is obtained by fitting the experimental mass of the 
$ \sum^{++}_{c} $ baryon (resulting in $ \mu=0.844$ fm$^{-1} $  ). 
Using $ \mu $   as a fit parameter instead of identifying it with the 
reduced mass $ m$ allows to make an accurate comparison between the
results of our present work and previous results
\cite{Ghalenovi:2011zz}. 

In Tables 3-8 the results for the masses and magnetic moments are 
presented and compared with other works 
\cite{Patel:2002jj, Ghalenovi:2012iu, Patel:2008nv, Albertus:2009ww,
  Majethiya:2008fe, Roberts:2007ni, Martynenko:2007je, Roberts:2007ni}
and experimental data 
\cite{Yao:2006px, Aaltonen:2007ar, Aaltonen:2007ap, Aaltonen:2009ny,
  Aaltonen:2011wd}. 
From Tables 3-8 we see that the results of the present work are in
good agreement with experimental data and other predictions. 
A comparison between the results of case I and the previous work 
\cite{Ghalenovi:2011zz} shows that the results obtained in case I 
are closer to experimental data. Note that the results of case II are 
very close to the ones obtained by Refs.\ \cite{Ghalenovi:2012zz, Ghalenovi:2012iu}. 

By comparing the results of cases I and II, we find that, apart from
the $ \Omega_{c} $ and $ \Omega_{b} $ baryons, the results obtained in
case II are overall closer to experimental data and other predictions 
than the ones obtained in case I and also in previous works. 
Also Tables 6-8 show that the predicted masses of double and 
triple heavy-flavor baryons in case II are closer to the results of other works.

\begin{table}
\begin{center}
Table 3.  Single charm baryon masses (masses are in MeV). The last two
  columns show the relative errors of cases I and II in comparison to
  experimental data. 
{\begin{tabular}{llllllllll} \hline  \hline  
Baryon & CaseI &CaseII& Exp. &Ref.\ \cite{Ghalenovi:2011zz} & Ref.\
\cite{Ghalenovi:2012zz} & Ref.\ \cite{Ghalenovi:2012iu} & Ref.\
\cite{Patel:2008nv} & Error I&Error II\\  \hline
$\sum_{c}^{++}$   &2454&2459&2454&2318&2452&2454&2425&0.0$ \% $&0.2$  \%$\\ 
$\sum_{c}^{++*}$  &2492&2508&2518&2446&2581&2526&2488&1.0$\%$&0.4$\%$ \\
$\sum_{c}^{0}$    &2459&2461&2453&2323&2457&2458&2442&0.2$\%  $&0.3$\%  $ \\
$\sum_{c}^{0*}$   &2497&2510&2518&2451&2586&2530&2507&0.8$\%  $&0.3$\%  $ \\
$\sum_{c}^{0}$    &2464&2462&2454&2328&2461&2460&2460&0.4$\%  $&0.3$\%  $\\
$\sum_{c}^{0*}$   &2503&2512&2518&2456&2591&2533&2526&0.6$\%  $&0.2$\%  $ \\
$\Xi_{c}^{+}$       &2576&2504&2468&2467&2466&2545&2512&4.4$\%  $&1.5$\%  $\\
$\Xi_{c}^{+*}$      &2634&2583&2647&2577&2596&2614&2584&0.5$\%  $&2.4$\%  $\\
$\Xi_{c}^{0}$    &2581&2506&2471&2453&Input&2547&2529&4.5$\%  $&1.4$\%  $\\
$\Xi_{c}^{0*}$     &2639&2585&2646&2582&2601&2616&2604&0.3$\%  $&2.3$\%  $\\
$\Omega_{c}^{0}$  &2715&2566&2697&2587&2476&2631&2601&0.7$\%  $&4.9$\%  $ \\
$\Omega_{c}^{0*}$ &2773&2648&2768&2716&2606&2700&2684&0.2$\%  $&4.3$\%  $\\ \hline
\end{tabular}}
\end{center}
\end{table}

\begin{table}
\begin{center}
Table 4. Single beauty baryon masses (masses are in MeV). 
The last two columns show the relative errors of cases I and II in 
comparison to experimental data.
{\begin{tabular}{l|l|l|l|l|l|l|l|l|l} \hline \hline  
Baryon & caseI & caseII &Exp. & Ref.\ \cite{Ghalenovi:2011zz} & Ref.\
\cite{Ghalenovi:2012zz} & Ref.\ \cite{Ghalenovi:2012iu} & Ref.\
\cite{Patel:2008nv}& Error I& Error II\\  \hline
$\sum_{b}^{+}$    &5834&5808&5807&5700&Input&5816&5772&$0.5\%$&$0.0\%$\\
$\sum_{b}^{+*}$    &5872&5858&5829&5826&5936&5888&5793&$0.7\%$&$0.5\%$ \\
$\sum_{b}^{0}$    &5839&5810&5811&-&-&5819&5793&$0.5\%$& $0.0\%$ \\
$\sum_{b}^{0*}$    &5877&5860&5832&-&-&5890&5816&$0.8\%$&$0.5\%$\\
$\sum_{b}^{-}$    &5844&5811&5815&5708&5818&5821&5816&$0.4\%$&$0.1\%$\\
$\sum_{b}^{-*}$    &5882&5861&5836&5836&5946&5892&5840&$0.8\%$&$0.4\%$\\
$\Xi_{b}^{0}$     &5956&5848&5787&5828&5821&5886&5880&$2.9\%$&$1.1\%$\\
$\Xi_{b}^{0*}$     &6014&5928&-&5957&5956&5972&5907&-&-\\
$\Xi_{b}^{-}$     &5961&5849&5792&5833&5826&5887&5903&$2.9\%$&$1.0\%$\\
$\Xi_{b}^{-*}$     &6019&5930&-&5962&5956&5974&5931&-&-\\
$\Omega_{b}^{-}$  &6095&5903&6054&5967&-&5986&5994&$0.7\%$&$2.5\%$\\
$\Omega_{b}^{-*}$  &6135&5986&-&6096&5961&6049&6028&-&-\\ \hline
\end{tabular}}
\end{center}
\end{table}

\begin{table}
\begin{center}
Table 5.  Magnetic moments of single charm and single beauty baryons 
with $ J^{p}=3/2^{+} $  in terms of the nuclear magneton $ \mu_{N} $ .\\
{\begin{tabular}{l|l|l|l|l|l|l} \hline \hline  
Baryon & caseI & caseII &Ref.\ \cite{Patel:2007gx} & Ref.\
\cite{Ghalenovi:2012iu} & Ref.\ \cite{Majethiya:2008fe} & Ref.\ \cite{Roberts:2007ni}\\  \hline
$\sum_{c}^{++*}$    &4.10&3.766&3.842&3.739&3.407&3.560\\
$\sum_{c}^{+*}$     &1.32&1.220&1.252&1.210&1.130&1.170 \\
$\sum_{c}^{0*}$     &-1.44&-1.333&-0.848&-1.322&-1.146&-1.230\\
$\Xi_{c}^{+*}$      &1.04&1.503&1.513&1.485&1.264&1.430\\
$\Xi_{c}^{0*}$      &-1.18&-1.124&-0.688&-1.111&-0.986&-1.000 \\
$\Omega_{c}^{0*}$   &-0.92&-0.903&-0.865&-0.887&-0.833&-0.770 \\
$\sum_{b}^{+*}$     &3.69&3.588&3.234&3.570&3.082&- \\
$\sum_{b}^{0*}$     &0.89&0.865&0.791&0.861&0.724&- \\
$\sum_{b}^{-*}$     &-1.91&-1.859&-1.655&-1.849&-1.634&- \\
$\Xi_{b}^{0*}$      &1.157&1.136&1.041&1.127&0.875&- \\
$\Xi_{b}^{-*}$      &-1.65&-1.621&-1.095&-1.609&-1.477&- \\
$\Omega_{b}^{-*}$   &-1.38&-1.380&-1.199&-1.365&-1.292&- \\ \hline
\end{tabular}}
\end{center}
\end{table}

\begin{table}

\begin{center}
Table 6. Double and triple charm baryon masses (in MeV).
{\begin{tabular}{l|l|l|l|l|l|l} \hline \hline  
Baryon & caseI & caseII &Ref.\ \cite{Ghalenovi:2011zz} & Ref.\
\cite{Ghalenovi:2012zz} & Ref.\ \cite{Martynenko:2007je} &Ref.\ \cite{Roberts:2007ni} \\  \hline
$\Xi_{cc}^{++}(ucc)$    &3703&3532&3579&3583&3510&3676\\
$\Xi_{cc}^{++*}(ucc)$   &3765&3623&3708&3722&3548&3753 \\
$\Omega_{cc}^{+}(scc)$   &3846&3667&3718&3592&3719&3815 \\
$\Omega_{cc}^{+*}(scc)$   &3904&3758&3847&3731&3746&3876 \\
$\Omega_{ccc}^{++*}(ccc)$   &5035&4880&4978&4842&4803&4965 \\ \hline
\end{tabular}}
\end{center}
\end{table}

\begin{table}
\begin{center}
Table 7. Double and triple charm baryon masses (in MeV).
{\begin{tabular}{l|l|l|l|l|l|l} \hline \hline  
Baryon & caseI & caseII &Ref.\ \cite{Ghalenovi:2011zz} & Ref.\
\cite{Ghalenovi:2012zz} & Ref.\ \cite{Martynenko:2007je} & Ref.\ \cite{Roberts:2007ni}\\  \hline
$\Xi_{bb}^{0}(ubb)$       &10467&10334&10339&10284&10130&10340\\
$\Xi_{bb}^{0*}(ubb)$      &10525&10431&10468&10427&10144&10367 \\
$\Omega_{bb}^{-}(sbb)$    &10606&10397&10478&10293&10422&10454 \\
$\Omega_{bb}^{-*}(sbb)$    &10664&10495&10607&10436&10432&10486 \\
$\Omega_{bbb}^{-*}(bbb)$   &15175&15023&15118&14810&14569&14834\\ \hline
\end{tabular}}
\end{center}
\end{table}

\begin{table}
\begin{center}
Table 8. Beauty and charm baryon masses (in MeV).
{\begin{tabular}{l|l|l|l|l|l|l} \hline \hline  
Baryon & caseI &caseII &Ref\cite{Ghalenovi:2011zz} & Ref.\
\cite{Ghalenovi:2012zz} & Ref.\ \cite{Albertus:2009ww} & Ref.\ \cite{Martynenko:2007je}\\  \hline
$\Omega_{cb}^{+}(ucb)$    &7087&6988&6959&6935&6928&6792 \\
$\Omega_{cb}^{+*}(ucb)$    &7145&7083&-&7076&-&6827 \\
$\Omega_{cb}^{0}(scb)$    &7226&7103&7098&6945&7013&6999 \\
$\Omega_{cb}^{0*}(scb)$   &7284&7200&-&7085&-&7024 \\
$\Omega_{ccb}^{+}(ccb)$   &8357&8190&8229&8038&-&8018 \\
$\Omega_{ccb}^{+*}(ccb)$  &8415&8290&8358&8186&-&8025 \\
$\Omega_{cbb}^{0}(cbb)$   &11737&11542&11609&11363&-&11280 \\
$\Omega_{cbb}^{0*}(cbb)$  &11795&11643&11738&11512&-&11287 \\ \hline
\end{tabular}}
\end{center}
\end{table}

\section{Summary }

In this paper we have studied masses and magnetic moments of
heavy-flavor baryons containing one, two, and three heavy-flavor
quarks in the ground-state $ (\gamma=0 ) $ for the different confining
potentials (1) and (2). Using the hypercentral approach we have
simplified the three-body problem and solved the Schr\"{o}dinger
equation numerically to obtain the ground-state energy eigenvalues and
eigenfunctions of baryonic systems. Hyperfine spin- and
isospin-dependent interactions result in small shifts of the baryon
energy.  Our results are similar to those of other works. The
confining interaction including a harmonic-oscillator term seems to 
give results closer to experimental data, especially for double and triple heavy baryons.
Our approach can also be used to study other three-body systems 
in the fields of nuclear, atomic, and molecular physics.


\begin{thebibliography}{0}
 
\bibitem{Albertus:2006ya}
  C.~Albertus, E.~Hernandez, J.~Nieves and J.~M.~Verde-Velasco,
  Eur.\ Phys.\ J.\ A {\bf 32} (2007) 183
   [Erratum-ibid.\ A {\bf 36} (2008) 119]
  
  
  
\bibitem{Ebert:2002ig}
  D.~Ebert, R.~N.~Faustov, V.~O.~Galkin and A.~P.~Martynenko,
  Phys.\ Rev.\ D {\bf 66} (2002) 014008
  
\bibitem{Zhang:2008pm}
  J.~-R.~Zhang and M.~-Q.~Huang,
  Phys.\ Rev.\ D {\bf 78} (2008) 094015

\bibitem{Guo:1996eg}
  X.~H.~Guo and T.~Muta,
  Mod.\ Phys.\ Lett.\ A {\bf 11} (1996) 1523
    
\bibitem{Ebert:2005xj}
  D.~Ebert, R.~N.~Faustov and V.~O.~Galkin,
  Phys.\ Rev.\ D {\bf 72} (2005) 034026
  
\bibitem{Albertus:2005uq}
  C.~Albertus, J.~E.~Amaro, E.~Hernandez and J.~Nieves,
  Nucl.\ Phys.\ A {\bf 755} (2005) 439
  
  
\bibitem{Flynn:2003vz}
  J.~M.~Flynn {\it et al.}  [UKQCD Collaboration],
  JHEP {\bf 0307} (2003) 066
  
\bibitem{Faessler:2006ft}
  A.~Faessler, T.~Gutsche, M.~A.~Ivanov, J.~G.~Korner, V.~E.~Lyubovitskij, D.~Nicmorus and K.~Pumsa-ard,
  Phys.\ Rev.\ D {\bf 73} (2006) 094013
  
\bibitem{Patel:2007gx}
  B.~Patel, A.~K.~Rai and P.~C. Vinodkumar,
  J.\ Phys.\ G {\bf 35} (2008) 065001
   [J.\ Phys.\ Conf.\ Ser.\  {\bf 110} (2008) 122010]

  
\bibitem{Patel:2008mv}
  B.~Patel, A.~Majethiya and P.~C.~Vinodkumar,
  Pramana {\bf 72} (2009) 679
  
\bibitem{Ghalenovi:2011zz}
  Z.~Ghalenovi, A.~A.~Rajabi and M.~Hamzavi,
  Acta Phys.\ Polon.\ B {\bf 42} (2011) 1849
  
  
\bibitem{Isgur:1978xj}
  N.~Isgur and G.~Karl,
  Phys.\ Rev.\ D {\bf 18} (1978) 4187
  
\bibitem{Capstick:1986bm}
  S.~Capstick and N.~Isgur,
  Phys.\ Rev.\ D {\bf 34} (1986) 2809
  
\bibitem{Giannini:1990pc}
  M.~M.~Giannini,
  Rept.\ Prog.\ Phys.\  {\bf 54} (1990) 453
  
\bibitem{Glozman:1995fu}
  L.~Y.~Glozman and D.~O.~Riska,
  Phys.\ Rept.\  {\bf 268} (1996) 263
  
\bibitem{Salehi:2011zza}
  N.~Salehi, A.~A.~Rajabi and Z.~Ghalenovi,
  Acta Phys.\ Polon.\ B {\bf 42} (2011) 1247
  
\bibitem{Hassanabadi:2008zz}
  H.~Hassanabadi, A.~A.~Rajabi and S.~Zarrinkamar,
  Mod.\ Phys.\ Lett.\ A {\bf 23} (2008) 527
  
  
\bibitem{Hassanabadi:2005z}
H.~Hassanabadi and A.~A.~Rajabi,
Few-Body\ Syst.\ {\bf 41} (2005) 201

  
\bibitem{Rajabi:2005}
A.~A.~Rajabi,
Few-Body\ Syst.\ {\bf 37} (2005) 4


\bibitem{Bali:2000vr}
  G.~S.~Bali {\it et al.}  [TXL and T(X)L Collaborations],
  Phys.\ Rev.\ D {\bf 62} (2000) 054503
  
\bibitem{Guillaume:2005v}
  G.~Plante and A.~F. ~Antippa,  
 J.\ Math.\ Phys. {\bf 46} (2005) 062108
  
  
\bibitem{Giannini:2003xx}
  M.~M.~Giannini, E.~Santopinto and A.~Vassallo,
  Prog.\ Part.\ Nucl.\ Phys.\  {\bf 50} (2003) 263
  
\bibitem{Ghalenovi:2012zz}
  Z.~Ghalenovi, A.~A.~Rajabi and A.~Tavakolinezhad,
  Int.\ J.\ Mod.\ Phys.\ E {\bf 21} (2012) 1250057
  
\bibitem{Patel:2002jj}
B.~Patel, A.~Majethiya and P.~C.~Vinodkumar,
  Phys.\ At.\ Nucl.\ {\bf 65} (2002) 917

 
\bibitem{Ghalenovi:2012iu}
  Z.~Ghalenovi and A.~Akbar Rajabi,
  Eur.\ Phys.\ J.\ Plus {\bf 127} (2012) 141
  
\bibitem{Patel:2008nv}
  B.~Patel, A.~K.~Rai and P.~C.~Vinodkumar,
  Pramana {\bf 70} (2008) 797
  
\bibitem{Albertus:2009ww}
  C.~Albertus, E.~Hernandez and J.~Nieves,
  Phys.\ Lett.\ B {\bf 683} (2010) 21
  
\bibitem{Majethiya:2008fe}
  A.~Majethiya, B.~Patel and P.~C.~Vinodkumar,
  Eur.\ Phys.\ J.\ A {\bf 38} (2008) 307
  
\bibitem{Roberts:2007ni}
  W.~Roberts and M.~Pervin,
  Int.\ J.\ Mod.\ Phys.\ A {\bf 23} (2008) 2817
  
\bibitem{Martynenko:2007je}
  A.~P.~Martynenko,
  Phys.\ Lett.\ B {\bf 663} (2008) 317
  
 
\bibitem{Yao:2006px}
  W.~M.~Yao {\it et al.}  [Particle Data Group Collaboration],
  J.\ Phys.\ G {\bf 33} (2006) 1
  
\bibitem{Aaltonen:2007ar}
  T.~Aaltonen {\it et al.}  [CDF Collaboration],
  Phys.\ Rev.\ Lett.\  {\bf 99} (2007) 202001
  
\bibitem{Aaltonen:2007ap}
  T.~Aaltonen {\it et al.}  [CDF Collaboration],
  Phys.\ Rev.\ Lett.\  {\bf 99} (2007) 052002
  
\bibitem{Aaltonen:2009ny}
  T.~Aaltonen {\it et al.}  [CDF Collaboration],
  Phys.\ Rev.\ D {\bf 80} (2009) 072003
  
\bibitem{Aaltonen:2011wd}
  T.~Aaltonen {\it et al.}  [CDF Collaboration],
  Phys.\ Rev.\ Lett.\  {\bf 107} (2011) 102001
  
\end{thebibliography}
\end{document}